\documentclass[aip,rsi,reprint,amsmath,amssymb,graphicx]{revtex4-1} 

\usepackage{graphicx}
\usepackage{dcolumn}
\usepackage{bm}
\usepackage[mathlines]{lineno}
\modulolinenumbers[5]
\linenumbers\relax 
\usepackage{color} 
\draft 

\begin{document}


\title{A contactless microwave-based diagnostic tool for high repetition rate laser systems} 



 \author{C. Braggio}
 \email[]{caterina.braggio@unipd.it}
 \affiliation{Dip. di Fisica e Astronomia and INFN sez. di Padova, Via F. Marzolo 8, I-35131 Padova, Italy}  \author{A. F. Borghesani}
 \affiliation{CNISM unit, Dip. di Fisica e Astronomia and INFN sez. di Padova, Via F. Marzolo 8, I-35131 Padova, Italy}


\date{\today}

\begin{abstract}
We report on a novel electro-optic device for the diagnostics of high repetition rate laser systems. It is composed of a microwave receiver and of a second order nonlinear crystal, whose 
irradiation with a train of short laser pulses produces a time-dependent polarization in the crystal itself as a consequence of optical rectification. 
This process gives rise to the emission of microwave radiation that is detected by a receiver and is analyzed to infer the repetition rate and intensity of the pulses. 
 We believe that this new method may overcome some of the limitations of photodetection techniques. 
\end{abstract}


\maketitle 

Mode-locked lasers with multi-gigahertz repetition rates are needed as key components in several applications \cite{Keller:2003}. They deliver pulses with duration ranging from picoseconds to a few femto\-se\-conds, with average output power up to several watts \cite{Krainer:2002}. 
The peak power can be so large to induce a nonlinear response in optical crystals, even without the use of additional optical amplification stages \cite{Innerhofer:2003}.

 Autocorrelation techniques \cite{Armstrong:1967,Sala:1980} are typically used to measure the single pulse duration, whereas the repetition rate of the pulses is determined by commercially available ultrafast photodiodes.
Their bandwidth is increased by reducing their active area and, as a result, their 
damage threshold is lowered while making them harder to align in absence of an input optical fiber.
 Besides, the high frequency response of the photodiodes is ultimately limited to a few GHz by long time tails in their output signals \cite{Gallo:2013}, although photodiodes have been built with bandwidth up to 100 GHz \cite{wang:1983b}.  
It is also worth noticing that laser oscillators have achieved repetition rates that cannot be detected by state-of-the-art photodetectors\cite{Krainer:2002}. 160 GHz repetition rate pulses, corresponding to 6 ps time separation, cannot be detected even by high-speed interdigitated detectors\cite{Gallo:2013}, whose best fall time is reported to be of the order of 30 ps. 

In the present work we describe a new diagnostic tool to characterize trains of high repetition rate laser pulses. 
The device is cost-effective and overcomes some limitations of state-of-the-art detection techniques. Moreover, in the transparency window of the nonlinear crystal used, for instance in KTP $350-4400$\,nm, its response is wavelength independent. 
Its application to a master oscillator power amplifier (MOPA) laser system is also described. 

\begin{figure}
\begin{center}
\includegraphics[width=\columnwidth]{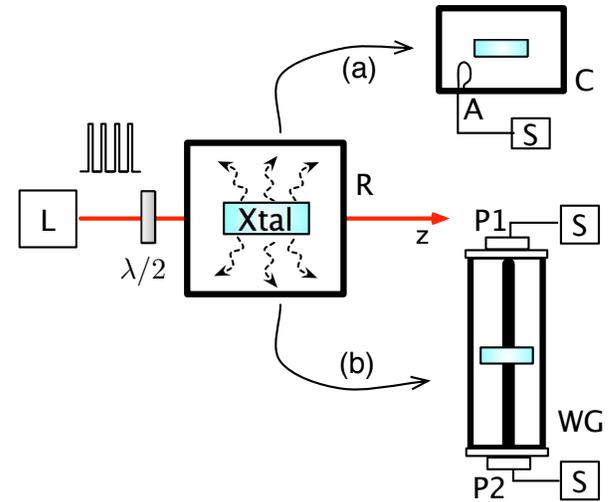}
\caption{Schematic diagram of the presented device. L denotes the laser system, Xtal the second-order nonlinear crystal, $\lambda/2$ the half-wavelength plate.
 R, the receiver, can be a cavity C as shown in option (a) or a waveguide WG as in (b). z is the laser propagation direction. The microwave signal generated by OR is detected by an antenna A in cavity C, or at ports P1 and P2 of a waveguide WG and fed to a sampling scope.}
\label{sc}
\end{center}
\end{figure}

The active element of the device is a nonlinear crystal mounted inside a microwave receiver R, either a cavity or a waveguide, as shown in Fig.~\ref{sc}. 
Laser pulses delivered by a mode-locked laser produce a time-dependent polarization in the crystal as a consequence of optical rectification (OR) \cite{bass1962}. 
This polarization is source of microwave radiation \cite{Borghe:2013} that can be transferred to any receivers without the bandwidth limitation of photodetection. 

In order to explain the device working principle, we use a simplified scalar model of the OR phenomenon, although the nonlinear electro-optic coefficients have tensor nature  \cite{yariv2007}. 
During each pulse of duration $\tau$ the laser electric field~$E(t)=E_0\cos \omega t$  induces in the crystal a nonlinear polarization:
\begin{equation}
\label{pnl}
P^{\mathrm{NL}}(t)=2 d E_0^2 \cos^2 \omega t =dE_0^2+dE_0^2\cos2\omega t
\end{equation}
where $d$ is the scalar second order nonlinear coefficient and $\omega$ is the angular frequency of the laser light. 
The last term in Eq.\,\ref{pnl} is responsible for second harmonic generation (SHG), whereas the first one, $P_0=dE_0^2$, is the result of OR. 
The pulse train delivered by a mode-locked laser has a frequency spectrum consisting of a series of discrete, regularly spaced lines \cite{Cundiff:2003}.  $P_0$ can also be expanded in Fourier series $P_0=dE_0^2\sum_m p_{m}\cos \left(m\omega_{\mathrm{RF}} t\right)$. $\omega_{RF}=2\pi\nu_{\mathrm{RF}}$, where $\nu_{\mathrm{RF}}$  is the repetition rate of the pulses.
Assuming that they are rectangular, the expansion coefficients are $p_m \propto \left(T/m\pi \right) \sin \left( \pi\tau/m T \right)$, in which a unitary phase factor has been dropped. $T=1/\nu_{\mathrm{RF}}$ is the time interval between the pulses.
The strength $E_{\mathrm{RF}}$ of the radiated microwave electric field  is proportional to the second time derivative of the polarization $\ddot P_0$ \cite{jackson}, 
\begin{equation}
\label{str}
E_{\mathrm{RF}}\propto \ddot P_0 =- dE_0^2 \omega_{\mathrm{RF}}^2 \sum\limits_m m^2  p_m \cos \left( m \omega_{\mathrm{RF}}t\right).
\end{equation}
Note that the radiated field is expected to be quadratic with $\omega_{\mathrm{RF}}$ so that the detection efficiency increases with increasing the laser repetition rate.

The emitted microwave radiation can be measured by placing the nonlinear crystal in either a cavity or a waveguide, as shown in Fig.~\ref{sc}. \textcolor{blue}{The laser beam to be characterized is allowed to impinge on the crystal by entering the receiver R through an aperture whose diameter is equal or slightly larger than the laser beam waist.  
} In both receivers, the amplitude of the microwave signal observed in the time domain   
is linearly proportional to $d$ and to the laser beam intensity $I \propto  E_0^2$. 
The cavity is designed to act as a tuned receiver at the first harmonic of the train whereas the waveguide configuration allows wideband detection of the generated radiation. Besides, the coaxial waveguide is a favorable geometry because it has no lower cutoff frequency and this allows detection also of low frequency instabilities in the laser intensity \cite{unp:2014}. 
\textcolor{blue}{Application of the present device to the detection of much higher repetition rates in pulsed laser systems could take advantage of the availability of commercial waveguides up to the D band ($90-180$\,GHz).}

The device has been tested with a MOPA system. It consists of (1) a CW Nd:YVO$_4$ oscillator that produces a continuous train of 12-ps pulses at 1064-nm wavelength with an adjustable repetition rate in the range $4.6 \, \mathrm{GHz}< \nu_{\mathrm{RF}}< 4.7\,$ GHz, (2) an acousto-optic modulator (AOM) that slices out a programmable duration train of optical pulses, and (3) two amplification stages \cite{Agnesi:08}.
The average output power of the mode-locked oscillator is 20 mW, whereas the single pulse energy of a few pJ can be amplified to 10 $\mu$J in trains of 500 ns duration. The beam area is set to $\approx 7$ mm$^2$ and the beam intensity has been varied during the present measurements from a few tenths up to several MW/cm$^{2}$.

The KTP (potassium titanyl phosphate, KTiOPO$_4$) crystal used (EKSMA Optics) is a parallelepiped  $4\times4\times10\,$mm$^3$ in size whose long axis is aligned in the direction $z$ of the laser propagation. 
An half wavelength plate is used \textcolor{blue} {in our measurements} to align the laser beam polarization along the direction of the largest second order electro-optical coefficient $d_{33}=15.7$ pm/V, which lies in the input face of the crystal.  
\textcolor{blue}{Note that the microwave signal can be made independent of the laser beam polarization, provided the nonlinear crystal is properly cut\cite{Borghe:2013}.
 }

 In order to minimize the impact of the device on the laser beam that passes through it, the crystal is cut in such a way not to satisfy the phase matching condition for SHG. In this way the efficiency of SHG is reduced below 1\,\%. 

\begin{figure}
\begin{center}
\includegraphics[width=\columnwidth]{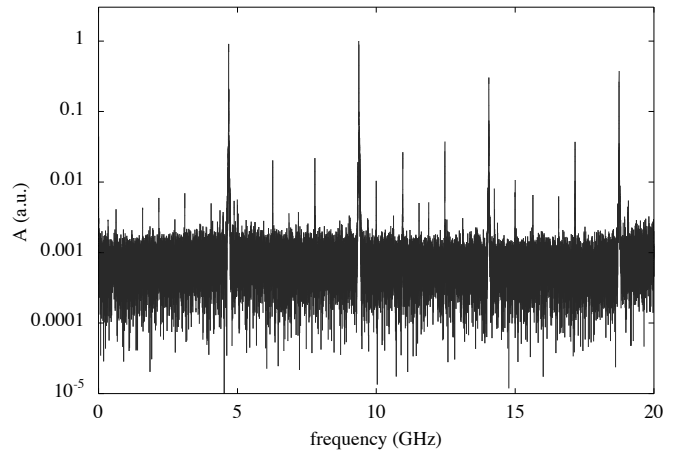}
\caption{\textcolor{blue}{Spectrum of the microwave signal generated by\,$\sim4.6$\,GHz repetition rate pulses impinging on a KTP crystal placed inside a coaxial waveguide as shown in Fig.~\ref{sc}(b). The spectrum extends up to the sampling frequency of the oscilloscope (LeCroy model SDA 820Zi-A, 20\,GHz oscilloscope).}  }
\label{20GHz}
\end{center}
\end{figure}

\textcolor{blue}{The spectrum of the microwave signal generated inside a coaxial waveguide receiver, is shown in  Fig.~\ref{20GHz}.}
\textcolor{blue}{As it is well known, the coaxial line supports higher order TE and TM waveguide modes in addition to the TEM mode. Our coaxial line has been designed so as to support the TEM mode up to approximately 13\,GHz, and observation of the first four microwave harmonics of the train of pulses (up to approximately 18.6\,GHz) was possible trough propagation of TE and TM modes.}

\begin{figure}[h!]
\begin{center}
\includegraphics[width=\columnwidth]{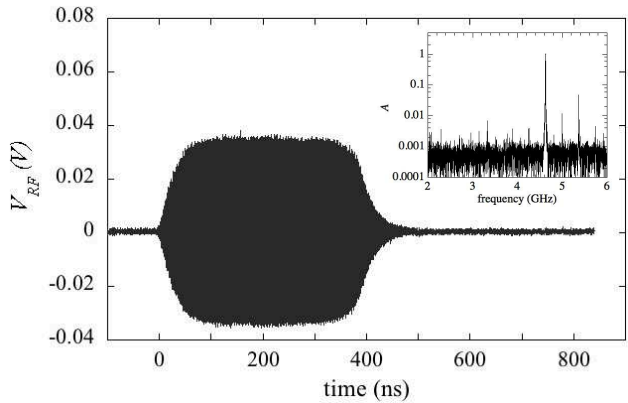}
\caption{Microwave signal detected in the cavity and its Fourier transform in the inset. A peak corresponding to the repetition rate of the pulses $\nu_{\mathrm{RF}}=4.662$ GHz is shown. The laser macro-pulse duration is 400 ns.}
\label{Rcav}
\end{center}
\end{figure}

\textcolor{blue}{As an example of narrow bandwidth detection}, we show in Fig.~\ref{Rcav} the microwave radiation received in a rectangular cavity, designed so as to sustain a TE$_{111}$ mode resonating at an adjustable frequency around 4.65\,GHz. 
In this way, the cavity selects the first harmonic $\nu_{\mathrm{RF}}$ of the train of pulses.
The emitted radiation excites the cavity mode and is detected by a critically coupled inductive loop, whose output is $V(t)=V_{\mathrm{RF}}\cos\omega_{\mathrm{RF}} t $.
The detected signal reaches 90\% of its stationary value in a time $\tau_{c}$ that 
is determined by the loaded quality factor of the cavity $Q_L \approx \pi \tau_c \nu_{\mathrm{RF}} $. The copper cavity used has
$Q_L \sim 1000$, yielding $\tau_c \geqslant 100$\,ns. No such rise time limitation is to be expected if a waveguide is used, instead.

To characterize the performance of our detector we define
a laser intensity-based figure of merit $F$ as:
\begin{equation}
\label{fmerito}
F=V_{\mathrm{RF}}/I \propto Gd\omega_{\mathrm{RF}}^2 \Omega 
\end{equation}
where $G$ is the overall microwave amplifier gain and $\Omega$ is the detector active volume, i.e., the smaller between the crystal volume $\Omega_c$ and the optical volume $\Omega_\mathrm{opt}$. 
In case of a gaussian beam, $\Omega_\mathrm{opt}=\pi w^2 l_\mathrm{opt}$, where $w$ is the beam waist and $l_\mathrm{opt}=c\tau/n$ is the optical length of the laser pulse, with $n$ index of refraction of the crystal. 
In our measurements $\Omega_\mathrm{opt}\approx 12$ mm$^3\ll\Omega_c=160\,$ mm$^3$. 

Using a 38\,dB gain microwave amplifier (Miteq, model AMF-4D-00112022-10P), signals corresponding to a laser intensity $\simeq 200$ kW/cm$^{2}$ were observed at the oscilloscope, corresponding to $F\simeq 5\cdot10^{-2}$ V/[MW/cm$^2$].
 Additional \textcolor{blue}{microwave} amplification stages can be employed to further decrease the detection threshold, provided the laser intensity $I$ is sufficient to excite the nonlinear response of the crystal. Therefore we expect that the present device can be used to monitor the CW laser emission from the recently developed mode-locked oscillators with multi-gigahertz repetition rates\cite{Keller:2003}.

We now describe an application specific for MOPA systems, in which only a finite number of pulses from a seed laser is amplified. The selected pulses envelope shape (macro-pulse) can be severely distorted during amplification due to gain saturation effects \cite{frantz:1963}. To compensate for this effect, the input drive signal of an acousto-optic device is properly shaped by a function generator \cite{braggio2011}. In order to accomplish this task it is crucial to have a detector that faithfully reproduces the pulse train. 

\begin{figure}
\begin{center}
\includegraphics[width=\columnwidth]{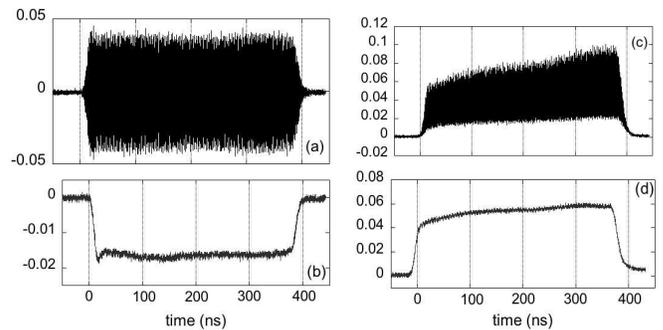}
\caption{(a) Microwave signal obtained in the wideband receiver configuration shown in part (b) of Figure~\ref{sc}. (b) Agilent 8474B 
planar-doped barrier diode detector output.
(c) Ultrafast photodiode  response (Advanced Laser Diode Systems, 11 ps rise time). (d) Response of the fast photodiode (DET10A/M Thorlabs Si biased detector, 1 ns rise time). The amplitude of all the signals is displayed in Volts.}
\label{4gr}
\end{center}
\end{figure}

 In Fig.~\ref{4gr} we show a comparison between the performance of the present device in the \textcolor{blue} {previously described coaxial waveguide configuration} and of commercially available fast and ultrafast photodetectors (PD). 
The fast PD has a time response of the order of 1\,ns, whereas we measured a fall time of 15\,ps in the case of  the ultrafast PD. The output pulse duration of the latter was measured using a Ti:Sa femtosecond laser (model Avesta TiF-50) whose repetition rate  is of the order of 80\,MHz. It is worth noticing that the ultrafast PD signal output voltage is not zero unless a time interval greater than 300\,ps is elapsed.
The plots in Fig.~\ref{4gr}\,(a)-(d) are recorded simultaneously. The intensity of each pulse is 200\,kW/cm$^2$.
To obtain the waveform shown in Fig.~\ref{4gr}(a), the microwave signal generated by the crystal in the waveguide is directly fed to a sampling oscilloscope (LeCroy WaveRunner 6000A, 6\,GHz bandwidth) connected to port P1 of the waveguide, as indicated in Fig.~\ref{sc}. A rectifying diode is connected to port P2 to get the signal shown in Fig.~\ref{4gr}(b). Commercial planar doped barrier diodes are available that have frequency response up to 50 GHz, and some diodes have been developed with response up to 94 GHz\cite{kearney:1991}. Therefore our device can potentially be used to monitor the envelopes of high repetition rate optical pulse trains without the use of expensive high-speed sampling oscilloscopes. 
The observed rise time of the signal, of the order of 30\,ns, is determined by the AOM bandwidth. No limitation in the device's response time is expected because of the electronic nature of the crystal response\cite{Borghe:2013}. 

The voltage waveform output of the ultrafast PD is shown in Fig.~\ref{4gr}(c). 
\textcolor{blue}{Since its response time is much smaller than the time interval between the laser pulses ($\sim 120$\,ps), its output displays the expected fine structure. However, we note that}  the voltage peaks increase with time, thereby indicating an apparent increase of the optical pulse intensity. This intensity increase is in contrast with the information given by the present device, as shown by the waveforms in Fig.~\ref{4gr}(a) and \ref{4gr}(b).
 This discrepancy might be explained by considering the long time tails in the optical pulse response of the photodiodes. This long tail degrades the ability of the photodiode to accurately monitor high repetition rate laser systems. Incidentally, the output from the fast PD shown in Fig.~\ref{4gr}(d) likewise shows a gradual increase in the height of the pulse envelope. 
    
Very recently the fall time of the output signal  has been reduced in high-speed photodetectors\cite{Gallo:2013}, but long time tails have not been shortened significantly. 
It is worth noticing that the fall time to 5\% of the peak amplitude is still comparable to 200 ps and a 5\% residual level might still be an issue when monitoring multi-GHz train of pulses. 
 
\begin{figure}[ht!]
\begin{center}
\includegraphics[width=\columnwidth]{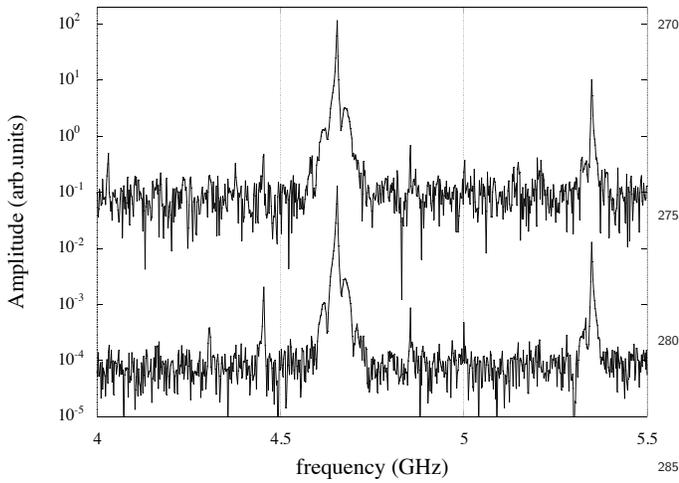}
\caption{Comparison between the FFT of the output signal from an ultrafast photodiode (top) and of the microwave emission signal in the coaxial waveguide (bottom). The ultrafast spectrum has been upwards shifted by 30\,dB for visibility purposes.}
\label{ffts}
\end{center}
\end{figure}

 The spectra of the signals displayed in (a) and (c) are compared in Fig.~\ref{ffts}. They are practically the same, thereby demonstrating that the present device can be equally well used as frequency counter in replacement of a more expensive ultrafast photodiode.
 \vglue 0.5 cm 

In conclusion, we have described a novel approach to the characterization of high repetition rate laser pulses that overcomes one of the limitations of state-of-the-art ultrafast photodiodes, i.e. the issues related to their long time tails in the detection of multi-GHz repetition rates. The present device is simple and cost-effective: only a commercially available nonlinear crystal and a receiver are necessary for its deployment. 
It is easy to use and does not require time-consuming alignment procedures. 
The device can be mounted in-path and without interfering with the optical beam properties of the laser output. It can act simultaneously as both an optical pulse repetition rate detector and, if properly aligned and calibrated, a laser intensity monitor. 
\textcolor{blue}{The definition of a figure of merit according to eq.~\ref{fmerito} allows intensity measurements from the amplitude of the microwave signal once an absolute measurement has been performed with a bolometer.}
The microwave radiation that is generated by the nonlinear crystal under the action of the laser pulses is independent of the optical wavelength and its responsivity extends to spectral regions not accessible by photodetection techniques. This characteristic is only limited by the transparency window of the nonlinear crystal chosen.

The authors wish to thank G. Carugno, G. Reali and G. Ruoso for valuable discussions and advice, F. Della Valle for critical reading of the manuscript.  We acknowledge financial support by INFN within the MIR experiment and thank E. Berto for technical assistance. 

\nocite{*}


\begin{thebibliography}{18}



\bibitem{Keller:2003}
U.~Keller.
\newblock {\em Nature}, 424(6950):831--838, 08 2003.

\bibitem{Krainer:2002}
L.~Krainer, R.~Paschotta, S.~Lecomte, M.~Moser, K.~Weingarten, and U.~Keller.
\newblock {\em IEEE J. Quant. Electron.}, 38:1331--1338, 2002.

\bibitem{Innerhofer:2003}
E.~Innerhofer, T.~S\"{u}dmeyer, F.~Brunner, R.~H\"{a}ring, A.~Aschwanden,
  R.~Paschotta, C.~H\"{o}nninger, M.~Kumkar, and U.~Keller.
\newblock {\em Opt. Lett.}, 28(5):367--369, Mar 2003.

\bibitem{Armstrong:1967}
J.~A. Armstrong.
\newblock {\em Appl. Phys. Lett.}, 10(1):16--18, 1967.

\bibitem{Sala:1980}
K.~Sala, G.~Kenney-Wallace, and G.~Hall.
\newblock {\em IEEE J. Quant. Electron.}, 16:990--996, 1980.

\bibitem{Gallo:2013}
E.~M. Gallo, A.~Cola, F.~Quaranta, and J.~E. Spanier.
\newblock {\em Appl. Phys. Lett.}, 102(16):161108, 2013.

\bibitem{wang:1983b}
S.~Y. Wang and D.~M. Bloom.
\newblock {\em Electron. Lett.}, 19(14):554--555, 1983.

\bibitem{bass1962}
M.~Bass, P.~A. Franken, J.~F. Ward, and G.~Weinreich.
\newblock {\em Phys. Rev. Lett.}, 9:446--448, 1962.

\bibitem{Borghe:2013}
A.~F. Borghesani, C.~Braggio, and G.~Carugno.
\newblock {\em Opt. Lett.}, 38, 4465--4468 (2013).

\bibitem{yariv2007}
A.~Yariv and P.~Yeh.
\newblock {\em {Photonics. Optical Electronics in Modern Communications}}.
\newblock Oxford University Press, Oxford, 2007.

\bibitem{Cundiff:2003}
S.~T. Cundiff and J.~Ye.
\newblock {\em Rev. Mod. Phys.}, 75(1):325--342, 03 2003.

\bibitem{jackson}
J.~D. Jackson.
\newblock {\em {Classical Electrodynamics}}.
\newblock Wiley, New York, 1998.

\bibitem{Agnesi:08}
A.~Agnesi, C.~Braggio, L.~Carr\`{a}, F.~Pirzio, S.~Lodo, G.~Messineo,
  D.~Scarpa, A.~Tomaselli, G.~Reali, and C.~Vacchi.
\newblock {\em Opt. Express}, 16(20):15811--15815, Sep 2008.

\bibitem{frantz:1963}
L.~M. Frantz and J.~S. Nodvik.
\newblock {\em J. Appl. Phys.}, 34(8):2346--2349, 1963.

\bibitem{braggio2011}
A.~Agnesi, C.~Braggio, G.~Carugno, F.~D. Valle, G.~Galeazzi, G.~Messineo,
  F.~Pirzio, G.~Reali, and G.~Ruo\-so.
\newblock {\em Rev. Sci. Instrum.}, 82:115107, 2011.

\bibitem{kearney:1991}
M.~J. Kearney, A.~Condie, and I.~Dale.
\newblock {\em Electron. Lett.}, 27:721--722, 1991.

\bibitem{wang:1983}
S.~Y. Wang, D.~M. Bloom, and D.~M. Collins.
\newblock {\em Appl. Phys. Lett.}, 42(2):190--192, 1983.

\bibitem{unp:2014}
A.~F. Borghesani, C.~Braggio, G.~Carugno, F.~Della Valle, G.~Ruoso.
\newblock {\em In preparation}.


\bibitem{Spuhler:2005}
G.~Spuhler, L.~Krainer, V.~Liverini, R.~Grange, M.~Haiml, S.~Pawlik,
  B.~Schmidt, S.~Schon, and U.~Keller.
\newblock {\em IEEE Photon. Tech. Lett.}, 17:1319--1321, 2005.






\end{thebibliography}
\end{document}